\documentclass[showpacs,preprint,prb,floatfix]{revtex4}
\usepackage{epsf}
\usepackage{psfig}
\usepackage{bm}
\usepackage{amssymb}
\begin{document}

%
%
\newcommand{\be}{\begin{equation}}
\newcommand{\ee}{\end{equation}}
\newcommand{\bea}{\begin{eqnarray}}
\newcommand{\eea}{\end{eqnarray}}
\newcommand{\beann}{\begin{eqnarray*}}
\newcommand{\eeann}{\end{eqnarray*}}
\newcommand{\bma}{\begin{array}{cc}}
\newcommand{\ema}{\end{array}}
\newcommand{\fr}{\frac}
\newcommand{\ra}{\rangle}
\newcommand{\la}{\langle}
\newcommand{\li}{\left}
\newcommand{\re}{\right}
\newcommand{\ri}{\right}
\newcommand{\uarr}{\uparrow}
\newcommand{\darr}{\downarrow}
\newcommand{\alp}{\alpha}
\newcommand{\df}{\stackrel{\rm def}{=}}
\newcommand{\nn}{\nonumber}
\newcommand{\dpl}{\displaystyle}
\def \gplus#1{{\cal G}^+ \left( #1 \right) }
\def \gminus#1{{\cal G}^- \left( #1 \right) }
\def \diff#1{{\cal D} \left( #1 \right) }
\def \lp {L_{\phi}}
\def \xo {x_1}
\def \xop {x_1^{\prime}}
\def \xt {x_2}
\def \yo {y_1}
\def \yop {y_1^{\prime}}
\def \yt {y_2}
%
%
\draft
\title{Aharonov-Bohm Physics with Spin II: \\
   Spin-Flip Effects in Two-dimensional Ballistic Systems }

\author{Diego Frustaglia}
\altaffiliation[Present address: ]{NEST-INFM \& Scuola Normale Superiore, 56126 Pisa, Italy.}
\affiliation{Institut f\"ur Theoretische
Festk\"orperphysik, Universit\"at Karlsruhe, 76128 Karlsruhe, Germany}
\author{Martina Hentschel}
\affiliation{
Duke University, Department of Physics, Box 90305, Durham, NC 27708-0305, U.S.A.}
\author{Klaus Richter}
\affiliation{ Institut f\"ur Theoretische Physik, Universit\"at
Regensburg, 93040 Regensburg, Germany}

%
\date{\today}
%
\begin{abstract}
{ We study spin effects in the magneto-conductance of ballistic mesoscopic
systems subject to inhomogeneous magnetic fields. We present a numerical
approach to the spin-dependent Landauer conductance which generalizes recursive
Green function techniques to the case with spin. Based on this method we address
spin-flip effects in quantum transport of spin-polarized and -unpolarized
electrons through quantum wires and various two-dimensional Aharonov-Bohm
geometries. In particular, we investigate the range of validity of a spin switch
mechanism recently found which allows for controlling spins indirectly via
Aharonov-Bohm fluxes. Our numerical results are compared to a transfer-matrix
model for one-dimensional ring structures presented in the first paper
(Hentschel {\it et al.}, submitted to Phys. Rev. B)
of this series. }
\end{abstract}

\maketitle
\bibliographystyle{simpl1}
%
%
\section{Introduction}


Topological quantum phases, like the Aharonov-Bohm (AB) phase \cite{AB59}
related to interference effects in 
the presence of a magnetic flux,
remain as a source of 
motivation
for the field of mesoscopic physics.
Another resource is the spin degree of
freedom that is responsible for a rich set of electronic transport
phenomena. 
Examples are manifold and range from
applications in the fast growing spintronics sector
\cite{DD90,P98,dassarma,KA99,WABDMRCT01,GB00}
to proposals for quantum information technology.
\cite{LD98,IABDLSS99,SSB03,SL03}
Experimental progress in the exploration of these phenomena relies on the
fabrication of very clean semiconductor heterostructures, the ability
to superimpose complex magnetic structures, and the development of
robust spin-injection mechanisms. \cite{spin-injection}

In mesoscopic quantum transport, 
novel spin-related
phenomena arise if the spin is coupled to non-uniform magnetic
fields.
This holds true
also
for Rashba (spin-orbit) effects,\cite{Rash60}
which can be regarded as arising from spin coupling to an intrinsic
effective magnetic field. 
In this context,
we expect signatures from
Berry\cite{Ber84} or geometric phases\cite{shaperewilcek},
topological quantum phases related to the change of the spin orientation upon
transport, to become accessible.
Recently a number of
experiments has been designed for directly manipulating the
spin dynamics via externally applied nonuniform magnetic
fields with amplitude or direction varying on mesoscopic
length scales\cite{YWGSKEN95,YTW98,DGNLMH00,NBH00}.  In this respect,
a ring geometry subject to such a textured magnetic
field is of particular interest since it represents a favorite
setup, at least in principle, for a direct observation of
Berry phases \cite{Ber84} in the magneto conductance\cite{Ste92,LSG93}.
The occurence of Berry phases, however, requires adiabatic spin transport
and therefore sufficiently strong inhomogenous magnetic fields.  Here
we will not address the recent discussion\cite{LKPB99,PFR03}
concerning the necessary conditions to be fulfilled for a clear-cut
measurement of such a geometrical phase.

Instead, we study
spin-dependent charge transport through ring-type conductors
in the entire range from weak to strong spin-magnetic field
coupling, i.e.\ from diabatic to adiabatic spin dynamics.
We will show that the interplay between the spatial and
spin part of the involved wavefunctions leads to subtle interference
effects which alter and enrich the usual AB picture of spinless particles.

After introducing our numerical approach to spin-dependent transport in the
Landauer framework in Sec.~\ref{NA}, we first illustrate in
Sec.~\ref{sec:strip}
basic spin features in the conductance for a simple strip geometry in a
nonuniform field. In Sec.~\ref{sec:adnonad} we investigate two-terminal
transport of unpolarized electrons through mesoscopic rings.
In Sec.\ \ref{sec:spinswitch}
we then examine in detail and generalize a recently proposed spin switch
mechanism in AB rings\cite{FHR01}. We demonstrate for a large class
of two-dimensional systems that this effect can be applied for controlling
indirectly (via an AB flux) the spin direction of spin-polarized particles
passing through mesoscopic ring geometries.

This is the second paper of a two-paper series on spin-dependent
Aharonov-Bohm physics. In a first paper \cite{HSFR03}, referred
to as paper I in the following, we gave an analytical proof of the
spin switch effect for one-dimensional rings. Here we compare our
different numerical results with analytical expressions derived in I.

\section{Numerical approach}
\label{NA}

\begin{figure}
\begin{center}
\centerline{ \psfig{figure=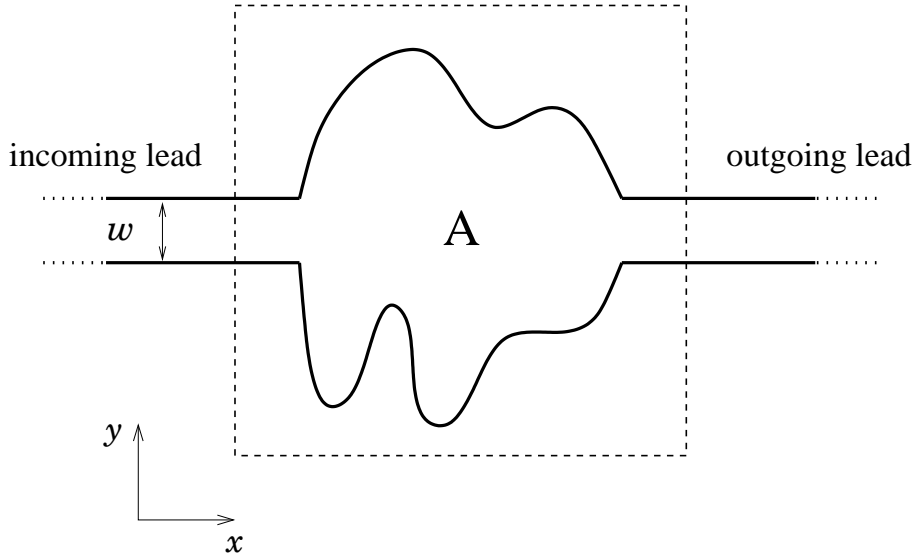,width=12cm,angle=0}}
\end{center}
\caption{
Two-dimensional conductor of arbitrary shape coupled to two semi-infinite
leads of width $w$ defining transverse modes (or channels,
with spin $s$ and spatial mode $m$) in the $y$-direction
due to confinement. Particles in channel ($s, m$) entering the region A from
the left are transmitted (reflected) to the right (left) after scattering into
channel ($s',n$).
}
\label{2D-structure}
\end{figure}

In this section we introduce the necessary concepts for studying
spin-dependent quantum transport in two-dimensional (2D) mesoscopic
systems
and outline our method used for the numerical calculation of the
magneto conductance.
The conductance of mesoscopic structures is commonly computed using the
Landauer formula \cite{Landa57,Dat97} for phase coherent transport in the
linear regime (small bias voltage). For leads of width $w$ supporting $N$ open
channels ($N=$Int$[k_{\rm F}w/\pi]$) the spin-dependent two-terminal
conductance at zero temperature reads
\be
\label{landa}
G=\frac{e^2}{h} \sum_{n,m=1}^N \sum_{s',s} T_{nm}^{s's}  \; .
\ee
Here, $T_{nm}^{s's}=|t_{nm}^{s's}|^2$ denotes the transmission
probability
between incoming ($m$) and outgoing ($n$) channels with
$s$ and $s'$ labeling the spin orientation
($s,s'=\pm 1$ denote  spin-up and -down states with respect to a
certain direction in space).
The transmission amplitude $t_{nm}^{s's}$ is obtained by projecting the
corresponding spin-dependent Green function ${\cal G}$ onto an
appropriate set of asymptotic
states (spinors) $\{\Phi\}$ defining incoming and outgoing channels
\cite{FL81}. For a 2D microstructure with
two attached leads as shown in Fig.~\ref{2D-structure},
the spin-dependent amplitudes read
\cite{FL81,SS88}
\be
\label{q-transm}
t_{nm}^{s's}=-i\hbar (v_n v_m)^{1/2} \int {\rm d}y' \int {\rm d}y
~\Phi_n^{s' *}(y')~{\cal G}(x_2,y';x_1,y;E)~\Phi_m^{s}(y) \; .
\ee
The above equation yields the transmission amplitude of an
electron with energy $E$ entering the cavity at $x=x_1$ in a transverse
spatial mode $m$ and spin state $s$, propagating through the
cavity by means of the Green function ${\cal G}$,
and escaping at $x=x_2$ in a mode $n$ with spin $s'$.
The quantities  $v_{n,m}$ denote the $x$-component of the particle
velocity in the modes $n$ and $m$, respectively.
Note that the spin-dependent Green function ${\cal G}$ is a $2
\times 2$ matrix.


The evaluation of the quantum meachanical transmission probability
(\ref{q-transm}) through
microstructures of
arbitrary shape (as shown in Fig.~\ref{2D-structure}) requires an
 efficient numerical tool for solving
the related transport equations. An adequate method is based on the use of a
tight-binding Hamiltonian, equivalent to a real-space discretization of the
Schr\"odinger equation, in combination with a recursive algorithm for computing
the corresponding Green function, see e.g.\ \cite{FG97,BDJS91} and
references therein. The advantages of the
method arise from its flexibility: different geometries (and topologies) are
readily handled, as well as the presence of magnetic fields or eventually disorder
potentials which can be easily included or modified. During the last decade
the method has nearly exclusively been used for the study of spin-independent
transport\cite{noteFR01}. In this article we generalize this
approach to account for spin-orbit scattering and for
the coupling of the spin degree of freedom to inhomogeneous
magnetic fields including non-adiabatic spin flip processes.


\begin{figure}
\begin{center}
\centerline{ \psfig{figure=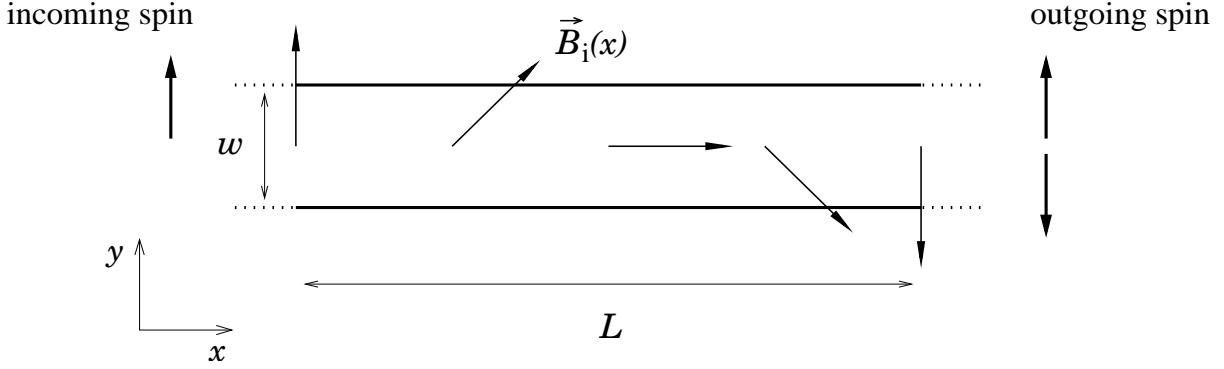,width=16cm,angle=0}}
\end{center}
\caption{
Two-dimensional straight ballistic conductor of width $w$ and length $L$
subject to a rotating in-plane magnetic field $\vec{B}_{\rm i}=B_{\rm
i}~[\sin[(\pi/L)x]~\hat{x} + \cos[(\pi/L)x]~\hat{y}]$, which vanishes
for $x < 0$ and $x > L$. The spin orientation of incoming and outgoing channels
is defined with respect to the $y$-axis.
}
\label{2D-lead}
\end{figure}


We consider non-interacting electrons with spin described by the
Pauli matrix vector $\vec{\sigma}$. The spin coupling to
a magnetic field $\vec{B}=\vec{\nabla}\times\vec{A}_{\rm
em}$ is accounted for by the Zeeman term
${\cal H}_{\rm s}=\mu\vec{B}\cdot\vec{\sigma}$.
Moreover, spin-dependent effects can arise in the absence of external
magnetic fields as well: The Rashba interaction ${\cal H}_{\rm
so}=\alpha_{\rm R}(\vec{\sigma} \times \vec{p})_{\rm z}/\hbar=i\alpha_{\rm
R}(\sigma_{\rm y}\partial/\partial x-\sigma_{\rm x}\partial/\partial y)$
\cite{Rash60} accounts for the spin-orbit coupling of strength $\alpha_{\rm R}$
in the presence of a vertical electric field (in $z$-direction).
The corresponding general Hamiltonian (for electrons of charge $-e$) then reads
\cite{noteRashba}
\bea
\label{spin-ham}
{\cal H} =  \fr{1}{2m^*}\li[\vec{p}+
    \fr{e}{c}\vec{A}_{\rm em}(\vec{r})\re]^2+ V(\vec{r}) + \mu~
\vec{B}(\vec{r})\cdot\vec{\sigma} + \frac {\alpha_{\rm
R}}{\hbar}\li[\vec{\sigma} \times
\li( \vec{p}+\fr{e}{c}\vec{A}_{\rm em}(\vec{r}) \re) \re]_{\rm z},~~~~~\;
\eea
where $\mu= g^* \mu_{\rm B}/2 =  g^* e \hbar/(4m_0\ c)$ is the magnetic moment,
$\mu_{\rm B}$ the Bohr magneton, $m_0$ the bare electron mass, $m^*$ the
effective electron mass, and $g^*$
the effective material-dependent gyromagnetic ratio.
The electrostatic potential $V(\vec{r})$ in Eq.~(\ref{spin-ham})
can represent, as in the present case,
the confining potential of a 2D ballistic conductor (see
Fig.~\ref{2D-structure}) or a disorder potential.

After introducing a 2D square grid of
spacing $a$ (identifying $x \equiv k a$ and $y \equiv l a$, with $k,l$
integers) we discretize the dimensionless Schr\"odinger equation
$(2m^*a^2/\hbar^2)(E-{\cal H})\Psi(\vec{r})=0$ corresponding to
Eq.~(\ref{spin-ham}) for spinors
\be
\label{spinor}
\Psi(\vec{r})={\Psi_1(\vec{r}) \choose \Psi_2(\vec{r})}.
\ee
Employing $\vec{p}=-i\hbar\vec{\nabla}$ and chosing the grid for a given field
strength $B$ such that $B a^2 \ll \phi_0 = hc / e$,
we arrive at the tight-binding representation \cite{thesis,MK-01}
\bea
\label{expan-sch-so}
\tilde{{\cal H}} &\equiv& \frac{2m^*a^2}{\hbar^2} (E-{\cal H})  \\
&=&\sum_{k,l}  \Bigg{\{}
\li ( \begin{array} {cc}
\tilde{h}_{kl}^{11} & \tilde{h}_{kl}^{12} \\
\tilde{h}_{kl}^{21} & \tilde{h}_{kl}^{22}
\end{array} \re )
|k,l \rangle \langle k,l |  + \nn \\
&+&\li[ \li( \begin{array} {cc}
\exp[i2\pi(aA_{\rm em}^{\rm x}/\phi_0)] & - \tilde{\alpha}_{\rm R}/2 \\
\tilde{\alpha}_{\rm R}/2 & \exp[i2\pi(aA_{\rm em}^{\rm x}/\phi_0)]
\end{array} \re )
|k,l \rangle \langle k+1,l | + {\rm h.c.}\re] + \nn \\
&+& \li[ \li ( \begin{array} {cc}
\exp[i2\pi(aA_{\rm em}^{\rm y}/\phi_0)] & i~\tilde{\alpha}_{\rm R}/2 \\
i~\tilde{\alpha}_{\rm R}/2 & \exp[i2\pi(aA_{\rm em}^{\rm y}/\phi_0)]
\end{array} \re )
|k,l \rangle \langle k,l+1 |
+ {\rm h.c.}\re] \Bigg{\}} \nn
\eea
with
\bea
\tilde{h}_{kl}^{11}&=& \tilde{E} -\tilde{V}-4-\tilde{\mu} B_{\rm z}\nn \; , \\
\tilde{h}_{kl}^{12}&=& - \tilde{\mu} (B_{\rm x}-iB_{\rm y}) -
\tilde{\alpha}_{\rm R} 2\pi a(A_{\rm em}^{\rm y}+iA_{\rm em}^{\rm
x})/\phi_0 \nn \; , \\
\tilde{h}_{kl}^{21}&=& -\tilde{\mu} (B_{\rm x}+iB_{\rm y}) -
\tilde{\alpha}_{\rm R} 2\pi a(A_{\rm em}^{\rm y}-iA_{\rm em}^{\rm x})/\phi_0
\nn \; , \\
\tilde{h}_{kl}^{22}&=& \tilde{E} -\tilde{V} -4+\tilde{\mu} B_{\rm z} \nn \; .
\eea
Here we introduced the scaled parameters $\tilde{\mu}=(2m^*a^2/\hbar^2) \mu$,
$\tilde{\alpha}_{\rm R}=(2m^*a/\hbar^2) \alpha_{\rm R}$,
$\tilde{E} =(2m^*a^2/\hbar^2)E$, and $\tilde{V} =(2m^*a^2/\hbar^2)V$.
Eq.~(\ref{expan-sch-so}) shows that the consideration of the
spin degree of freedom gives rise to generalized on-site and hopping energies
represented by $2 \times 2$ matrices. The relative magnitude of
the matrix elements determines, besides aspects related to the orbital
motion, the spin dynamics of the carriers, generally leading to spin flips.
We note that the Zeeman interaction $\tilde{\mu} B_{\rm z}$ does not enter into
the hopping terms in Eq.~(\ref{expan-sch-so})
because the Zeeman term does not involve derivatives.
For vanishing spin-orbit coupling ($\tilde{\alpha}_{\rm R}=0$ in
Eq.~(\ref{expan-sch-so})) the hopping matrices are diagonal in spin space
and only the on-site terms generate spin flips.

For the calculation of the spin-dependent Green function matrix ${\cal G}$
we generalized the method used for spinless particles \cite{BDJS91}.
This is based on the Dyson equation,
\be
\label{dyson-eq}
{\cal G}={\cal G}_0+{\cal G}_0 U {\cal G}={\cal G}_0+{\cal G} U {\cal G}_0 \:,
\ee
which relates the Green function ${\cal G}_0$ of the unperturbed system to
the Green function ${\cal G}$
of the system under  perturbation $U$.
In the present case $U$ is given by the hopping terms
$|k,l \rangle \langle k+1,l |$ and $|k+1,l \rangle \langle k,l |$
in Eq.~(\ref{expan-sch-so}).
Relation (\ref{dyson-eq}) represents an implicit equation for ${\cal G}$,
leading to a system of equations which can be solved recursively
\cite{FG97,thesis}.
A similar method has recently been used for the study
of the spin-orbit coupling in quasi-ballistic and disordered
wires \cite{NF01,PB02}.

While the approach outlined above is rather general and applicable to spin
transport in a variety of different systems, from now on
we focus on pure Zeeman coupling in the presence of inhomogeneous magnetic
fields, i.e.\
$\alpha_{\rm R}\equiv 0$ in the Hamiltonian
(\ref{spin-ham}). In addition, we consider ballistic (disorder free) dynamics.
Apart from this purely numerical approach, corresponding analytical solutions
have been obtained in the limit of 1D rings in paper I and \cite{Ste92}.
We compare both results in Sec.~\ref{sec1dpart}.

%
\section{Case study: quantum wires}
\label{sec:strip}

To illustrate our spin-dependent numerical approach and basic spin effects
we consider first the simple geometry
shown in Fig.~\ref{2D-lead}. It consists of a straight ballistic conductor of
width $w$ and length $L$ in a rotating in-plane magnetic field of the form
$\vec{B}_{\rm i}=B_{\rm
i}~[\sin[(\pi/L)x]~\hat{x} + \cos[(\pi/L)x]~\hat{y}]$, which vanishes at
leads for $x < 0$ and $x > L$. Chosing
a proper gauge, the $x-$ and $y-$components of the vector potential
$\vec{A}_{\rm em}$ generating the field $\vec{B}_{\rm i}$ vanish in the plane
$z=0$ \cite{noteGauge}. Moreover, as depicted in Fig.~\ref{2D-lead}, we consider
asymptotic spin states defined with respect to the $y$-axis,
namely, eigenvectors of $\sigma_{\rm y}$.


\begin{figure}
\begin{center}
\centerline{ \psfig{figure=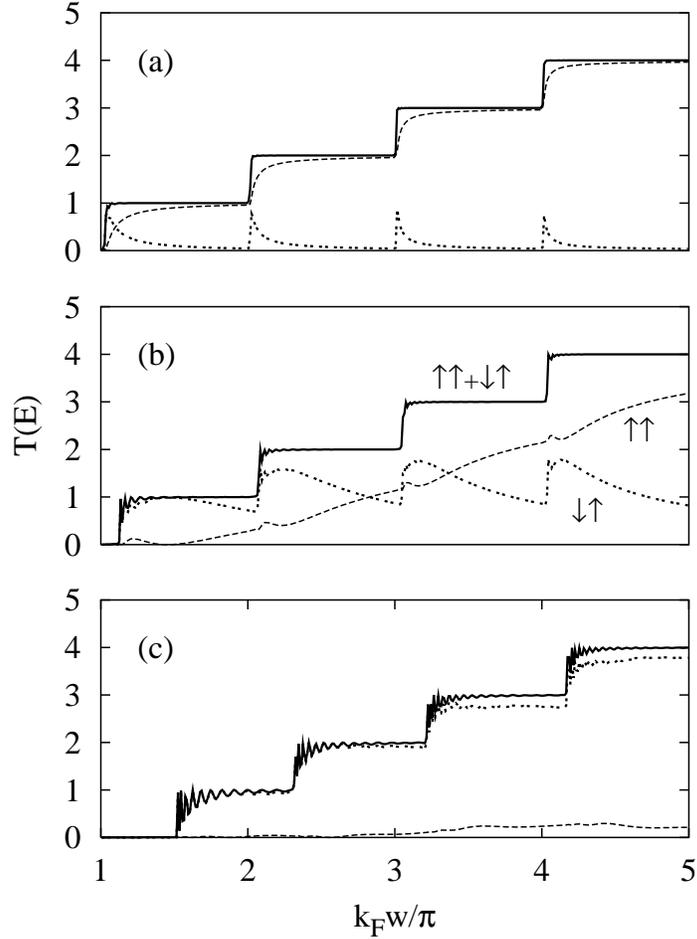,width=10cm,angle=0}}
\end{center}
\vspace*{-5mm}

\caption{
Numerically calculated spin-dependent transmission through the straight
2D conductor of Fig.~\ref{2D-lead} using the generalized
recursive Green function method introduced in Sec.~\ref{NA}. Results
are shown for the quantum transmission of incoming spin-up polarized
particles ($\uparrow$, defined with respect to
 the $y$-axis, see Fig.~\ref{2D-lead}) in the presence of a weak (a),
moderate (b), and strong (c) magnetic field as a function of the dimensionless
wave number $k_{\rm F}w/\pi$. The overall transmission (solid line) is split
into its components $T^{\uparrow \uparrow}$ (dashed line) and $T^{\downarrow
\uparrow}$ (dotted line) corresponding to outgoing spin-up ($\uparrow$) and
-down ($\downarrow$) channels, respectively.
}
\label{T-strip-up}
\end{figure}


Our numerical results are
summarized in Fig.~\ref{T-strip-up}. It shows the quantum transmission of incoming
spin-up electrons ($\uparrow$) in the presence of a weak (a), moderate (b),
and strong (c) magnetic field as a function of the dimensionless wave number
$k_{\rm F}w/\pi$. The overall transmission (solid line) is decomposed into its
components $T^{\uparrow \uparrow}$ (dashed line) and $T^{\downarrow \uparrow}$
(dotted line) corresponding to outgoing up- ($\uparrow$) and down-polarized
($\downarrow$) spin channels, respectively.

For spinless transport, the conductance
through the 2D wire of Fig.~\ref{2D-lead} is quantized
as for a quantum point contact \cite{WHBWKMF88}.
The conductance exhibits steps of size $\Delta G=  2 e^2/h$
equivalent to steps $\Delta T=1$ in the transmission, each time a new
transverse
mode is opened in the conductor. In the presence of the magnetic field
$\vec{B}_{\rm i}$, the overall transmission $T^{\uparrow \uparrow}
+ T^{\downarrow \uparrow}$ (solid line in Fig.~\ref{T-strip-up}(a-c))
shows the same steps, up to small deviations for
strong $B_{\rm i}$. However, the individual spin components vary considerably
as the field strength  $B_{\rm i}$ increases (from (a) to (c)).
The relative strength of the field $B_{\rm i}$
can be characterized by comparing the relevant
time scales in the system \cite{Ste92,LSG93,LKPB99,PFR03,FHR01,FR01,thesis},
i.e.\
the Larmor frequency  $\omega_{\rm s}=2\mu B_{\rm i}/\hbar$
of spin precession around the local field with the characteristic
frequency $\omega \sim v/L$ of orbital motion (with velocity $v$)
in the region where the direction of the field changes significatively
\cite{noteSLF}.

In the weak field limit ($\omega \gg \omega_{\rm s}$) the
spin dynamics is slow compared to the orbital motion.
In this situation the {\em non-adiabatic}
channel $T^{\uparrow \uparrow}$ (dashed line in Fig.~\ref{T-strip-up}(a)) 
dominates the transport, and most electrons leave
the conductor conserving the incoming
spin-up polarization.
The opposite limit of a strong field is given when $\omega \ll \omega_{\rm s}$.
Here the spin stays {\it adiabatically} aligned with the spatially varying
magnetic field direction during transport, such that the spin orientation
is finally reversed at the outgoing lead:
the particles escape
in spin-down state after travelling through the conductor, and the main
contribution to the transmission is given by $T^{\downarrow \uparrow}$
(dotted line in Fig.~\ref{T-strip-up}(c)).
We thus refer to $T^{\downarrow \uparrow}$ and $T^{\uparrow \uparrow}$
as the adiabatic and  non-adiabatic transmission channels, respectively.
For the intermediate case of moderate
fields, $\omega \sim \omega_{\rm s}$, Fig.~\ref{T-strip-up}(b), the
contribution of both adiabatic and non-adiabatic channels is comparable.
Moreover, there is further structure in the quantum transmission which we
discuss in the following:

(i) The adiabatic channel $T^{\downarrow \uparrow}$ is enhanced
each time a new transverse mode opens, see dotted lines
in Fig.~\ref{T-strip-up}(a,b). This is related to the fact that
different modes propagate with different velocities
$v_x(n)$ along the $x$-direction, and the mode specific orbital
frequency $\omega=\omega_n=v_x(n)/L$ has to be compared with $\omega_{\rm s}$
in order to determine whether the correponding spin propagates adiabatically
or not \cite{noteModes}.
When a new mode just opens, the associated particle velocity $v_x$ vanishes,
leading to a small $\omega$ and thereby giving rise to
a large adiabatic contribution.

(ii)
The staircase profile of the overall transmission in Fig.~\ref{T-strip-up} is
shifted to higher values of $k_Fw/\pi$ (corresponding to higher Fermi energies
for the incoming electrons) as $B_{\rm i}$ increases 
from  Figs.~\ref{T-strip-up}(a) to (c).
This is due to a Zeeman {\it barrier} which the incoming spin-up electrons
must overcome in  the adiabatic limit of strong field.
 Incoming  spin-down electrons experience a
Zeeman {\it well} which does not lead to such a shift.
(In Fig.~\ref{T-strip-up} only spin-up polarized electrons are shown.) 

(iii)
We observe an oscillatory pattern modulating the plateaus
in the overall transmission (solid line) as the adiabatic limit is
approached (Fig.~\ref{T-strip-up}(c) in particular).
This is also a consequence of the Zeeman
barrier, giving rise to interfering backscattered waves, similar to the
textbook case of a  wave scattered at a 1D rectangular potential barrier.

We finally point out that for the simple geometry of Fig.~\ref{2D-lead} Berry
phases \cite{Ber84} arising in the adiabatic limit do not play a role
in the conductance because their observation 
requires doubly-connected
geometries
\cite{Ste92,FR01,LGB90}.
However, the introduction of disorder could lead to signatures of Berry phases
in the magneto conductance via the suppression of weak-localization
\cite{KTO03}.

%
\section{From diabatic to adiabatic spin transport in mesoscopic rings}
\label{sec:adnonad}

The above discussion of spin transport through a straight wire 
geometry
illustrated basic spin-dependent transport phenomena. Here we
discuss  more sophisticated spin-dependent effects arising in
transport of unpolarized electrons through mesoscopic ring geometries
subject to different field textures.

\subsection{Magnetic field setup}

\begin{figure}
\begin{center}
\centerline{\psfig{figure=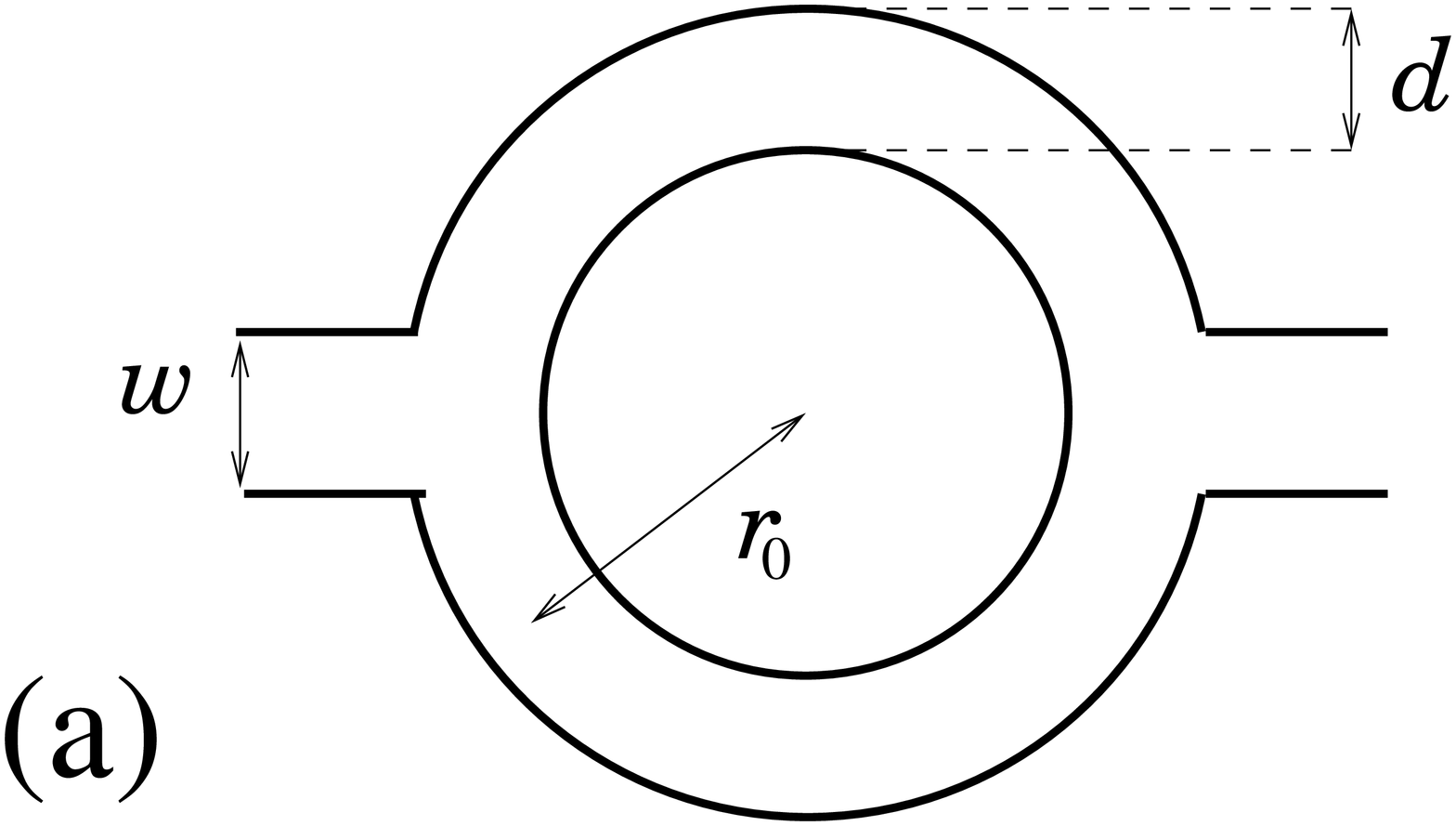,width=7cm,angle=-90}
\psfig{figure=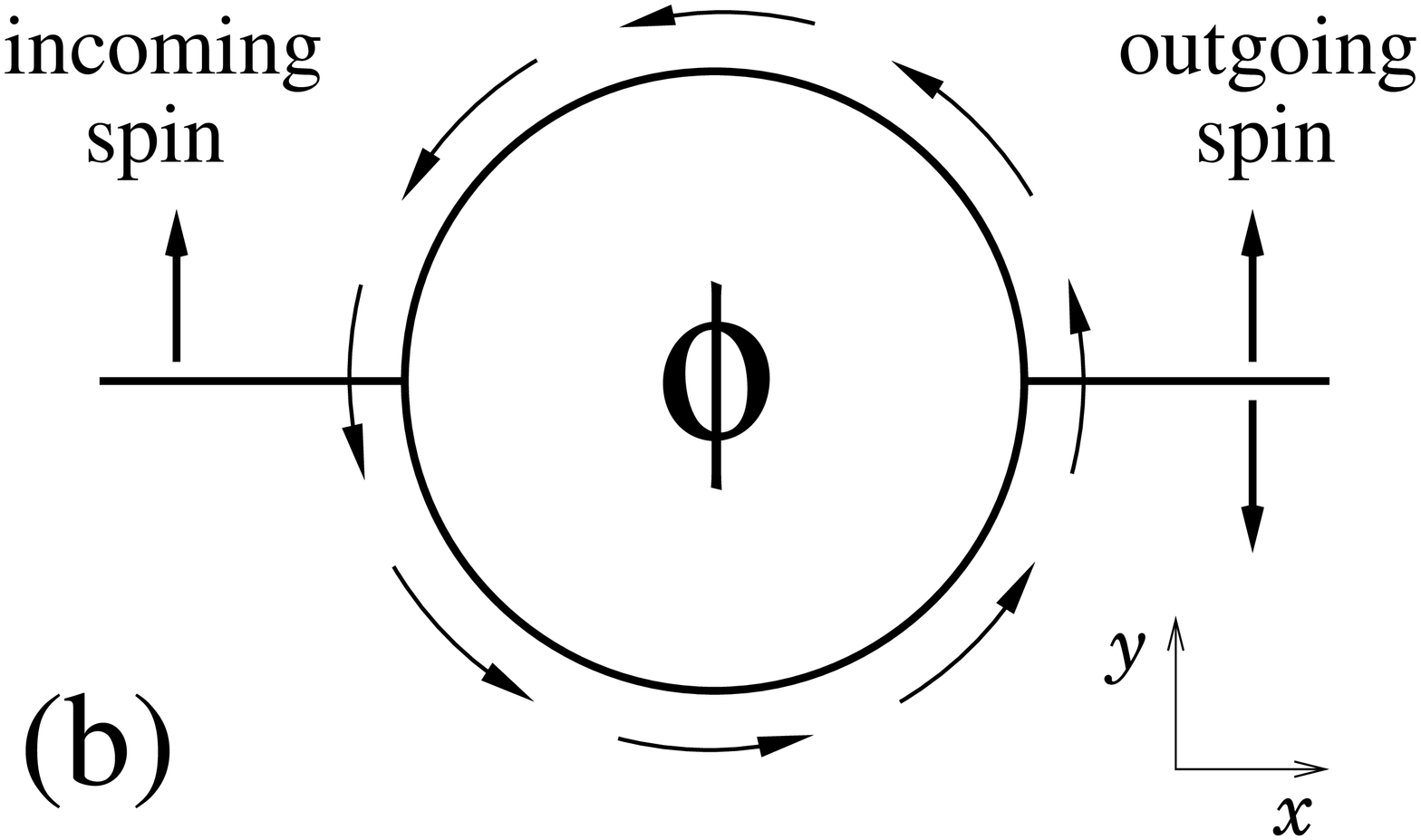,width=7cm,angle=-90}}
\end{center}
\caption{
Geometries of ballistic microstructures used in the
quantum calculations of the spin-dependent conductance for
a circular (in-plane) magnetic field texture $\vec{B}_{\rm i}$,
Eq.~(\ref{Bi}), plus a magnetic flux $\phi$.
Spin directions are defined with respect to the $y$-axis.
}
\label{Q-1D-rings}
\end{figure}

\begin{figure}
\begin{center}
\centerline{ \psfig{figure=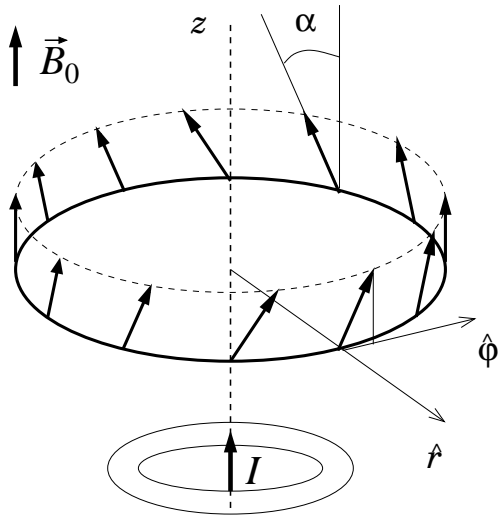,width=10cm,angle=-90}}
\end{center}
\vspace*{-5mm}

\caption{
Magnetic field texture corresponding to a circular field generated by an
electric current $I$ along the $z$-axis (Eq.~(\ref{Bi})) plus an additional
uniform field $\vec{B}_0=B_0~\hat{z}$, defining a tilt angle $\alpha$. The
field is evaluated on a one-dimensional ring contour.
}
\label{crown-field}
\end{figure}

We consider  ballistic ring structures with two attached
leads (see  Fig.~\ref{Q-1D-rings}) subject to a magnetic field which
has two contributions, $\vec{B}(\vec{r})=\vec{B}_0+\vec{B}_{\rm i}(\vec{r})$.
The first term corresponds to a perpendicular uniform magnetic field
$\vec{B}_0=B_0 \hat{z}$ leading to a magnetic flux $\phi=\pi r_0^2 B_0$
(where $r_0$ is the mean radius of the ring) to be used as a
tunable parameter to study the magneto conductance of the microstructures.
The second contribution to $\vec{B}(\vec{r})$ is given by
circular in-plane
field (see Fig.\ref{Q-1D-rings}(b))  reading in polar coordinates
\be
\label{Bi}
\vec{B}_{\rm i}(\vec{r}) = B_{\rm i}(r)~\hat{\varphi} =  \frac {a}{r}~\hat{\varphi}.
\ee
Such a field can be viewed as being generated  by a perpendicular
electrical current $I$ through the center of the microstructure
with $a=\mu^\ast I/2\pi$
\cite{noteKABMR00}. The configuration of the overall magnetic field is
schematicaly represented in 
Fig.~\ref{crown-field}.


%

\subsection{Comparison between analytical and numerical results}
\label{sec1dpart}

We begin our analysis of the spin-dependent conductance through mesoscopic rings
by comparing numerical results for 2D rings, based on the technique described in
Sec.\ \ref{NA}, with results 
for one-dimensional (1D) rings derived in paper I \cite{HSFR03}. 
To this end, we
consider the spin-dependent transmission of unpolarized electrons
in the entire crossover regime between zero magnetic field (diabatic limit,
$B_{\rm i}\! =\! 0$) and the adiabatic limit ($B_{\rm i}$ large).
We quantify the degree of adiabaticity
in terms of the parameter
\begin{equation}\label{eq:Q}
Q\equiv \omega_s / \omega \; ,
\end{equation}
the ratio between 
the spin precession frequency $\omega_s$
and the orbital frequency $\omega$.
Hence, $Q$ increases as adiabaticity is approached.


\begin{figure}
\begin{center}
\centerline{\psfig{figure=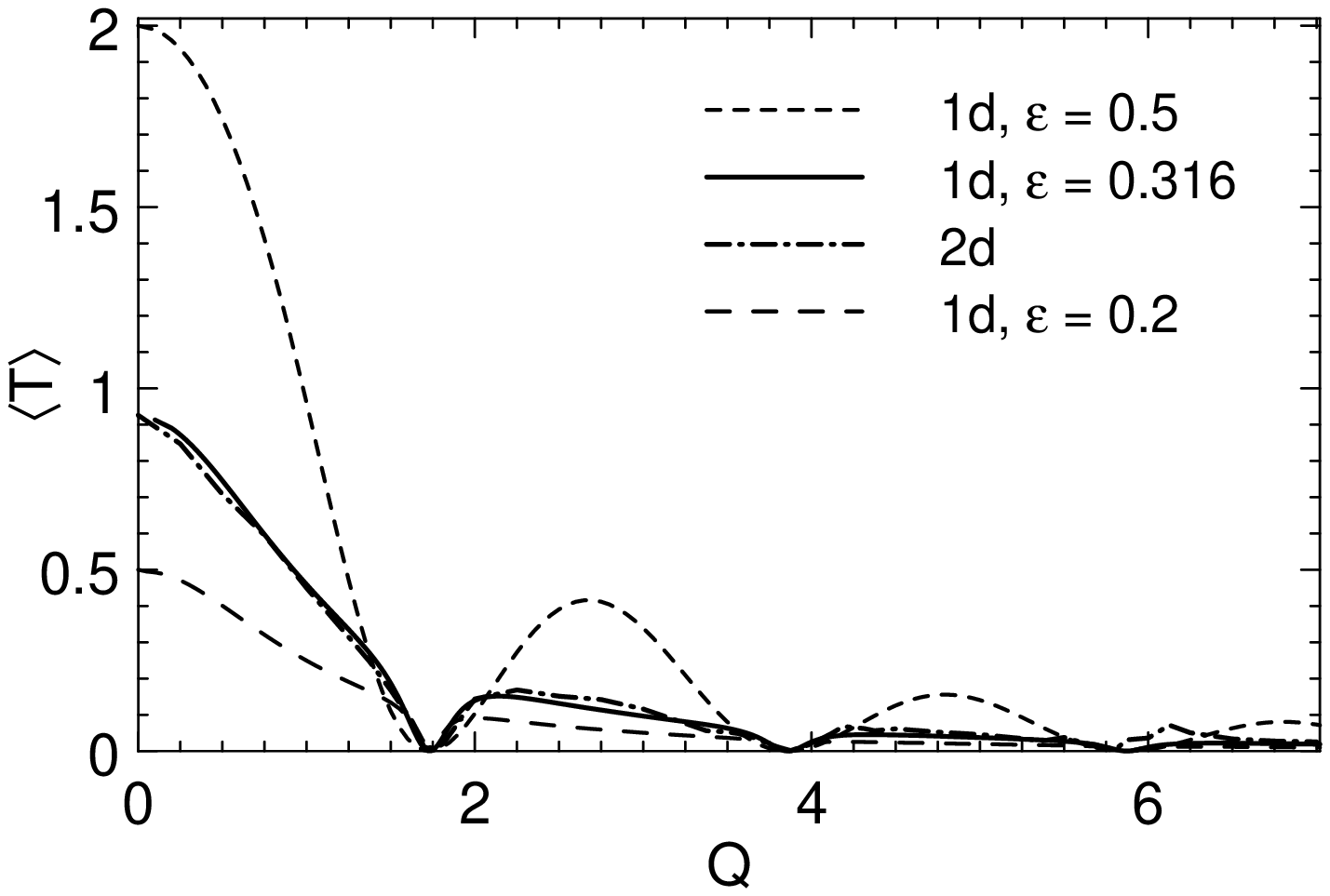,width=14cm,angle=0}}
\end{center}
\vspace*{-5mm}
\caption{
Spin-dependent quantum transmission of unpolarized
electrons through a 1D ring calculated within the transfer
matrix approach of paper I. 
The inhomogeneous magnetic field lies in the
plane of the ring, see Fig.~\ref{Q-1D-rings}.
The energy-averaged transmission
$\langle T \rangle$ as a function of the adiabaticity
parameter $Q$ is shown 
for coupling parameters $\epsilon=0.2,\; 0.316,\; 0.5$ and $\phi = 0$.
The numerical result of the 2D tight-binding approach
for a ring with one open channel is given by the dashed-dotted line
and follows closely the curve for $\epsilon=0.316$.
}
\label{epsilon-dep-T}
\end{figure}

The numerical quantum mechanical result for the energy-averaged conductance in a
quasi 1D ring, i.e. a 2D ring with just one open channel, as a function of $Q$
is shown as the dashed-dotted line in Fig.~\ref{epsilon-dep-T}. It
exhibits an overall Lorentzian decay  \cite{FHR01} and distinct zeros for
certain field strengths.

These features can be well understood within a complementary 1D model
built from a  (strictly) 1D ballistic ring coupled to 1D leads.
In this case, the Schr\"odinger equation for the
Hamiltonian (\ref{spin-ham}) can be solved exactly
for all values of the adiabaticity parameter $Q$ (if $V(\vec{r})=0$),
see paper I \cite{HSFR03} of this series for
details and references.
In short, the 1D model in paper I employes a transfer matrix approach
to describe transport properties,
making use of the exact eigenstates of the 1D Hamiltonian.
The coupling between the leads and the
1D ring is specified by a scattering matrix
with  coupling constant $\epsilon$ as parameter.
Zero and strongest coupling is described by $\epsilon=0$ and
$\epsilon=0.5$, respectively.


As an example we show in Fig.~\ref{epsilon-dep-T} results for the
energy-averaged transmission probability $\la T(Q,\epsilon) \ra $
in the case of a circular in-plane magnetic field ($\alpha=\pi/2$) with
Aharonov-Bohm flux $\phi = 0$.
The transmission
vanishes in the adiabatic limit ($Q
\to \infty$) due to the presence of a Berry phase that leads to an additional
phase shift. Its action can be interpreted
as a geometric flux\cite{FR01} of half a flux quantum acting similar to an
Aharonov-Bohm flux. Hence, for $\phi=0$ the additional
Berry phase causes destructive interference of the waves in the two arms of
the ring, leading to a vanishing transmission. 
We refer the interested reader to the detailed discussion in paper I. 

In paper I it is also shown that the energy-averaged transmission
vanishes due to destructive interference at points 
$Q=\sqrt{4m^2-1}$ with integer $m$, 
i.e. $Q = \sqrt{3}, \sqrt{15}, \ldots $, in
agreement with the zeros observed in Fig.~\ref{epsilon-dep-T}.
This gives rise to the observed oscillating structure of the averaged
transmission
probability $\la T \ra$ and holds for all coupling strengths $\epsilon$,
see Fig.~7 in paper I.  Hence, tuning $Q$ (i.e.\
the strength of the inhomogeneous field) enables for controlling
the overall conductance via the coupling of the electron spin to the field.

While the coupling parameter $\epsilon$ appears naturally in
the analytical transfer matrix approach in paper I, 
an effective $\epsilon$ is
difficult to determine
from the 2D tight-binding calculations described in Sec.~\ref{NA}.
We have used the quasi-1D tight-binding transmission at $Q$=0 to fix
the parameter $\epsilon$ to 0.316 in the transfer matrix approach.
This choice yields considerable agreement between the quasi-1D numerical
and 1D analytical curves in the whole $Q$-regime, see 
Fig.~\ref{epsilon-dep-T}.
We further note that in an experimental setup the effective coupling
$\epsilon$ could
be controlled by means of local gates acting as tunable potential barriers
at the junctions.

%
\section{Magneto conductance of spin-polarized currents}
\label{sec:spinswitch}

In the following we show how the transmission and polarization of
{\em spin-polarized} electrons in various ring geometries can be affected by
an additional magnetic flux $\phi = \pi r_0^2 B_0$.
Numerical results for single and multichannel
transport are presented and discussed. Different ring geometries and field
textures are considered.

%

\subsection{Aharonov-Bohm ring as a spin switch}
  \label{ABRSS}

\begin{figure}
\begin{center}
\centerline{ \psfig{figure=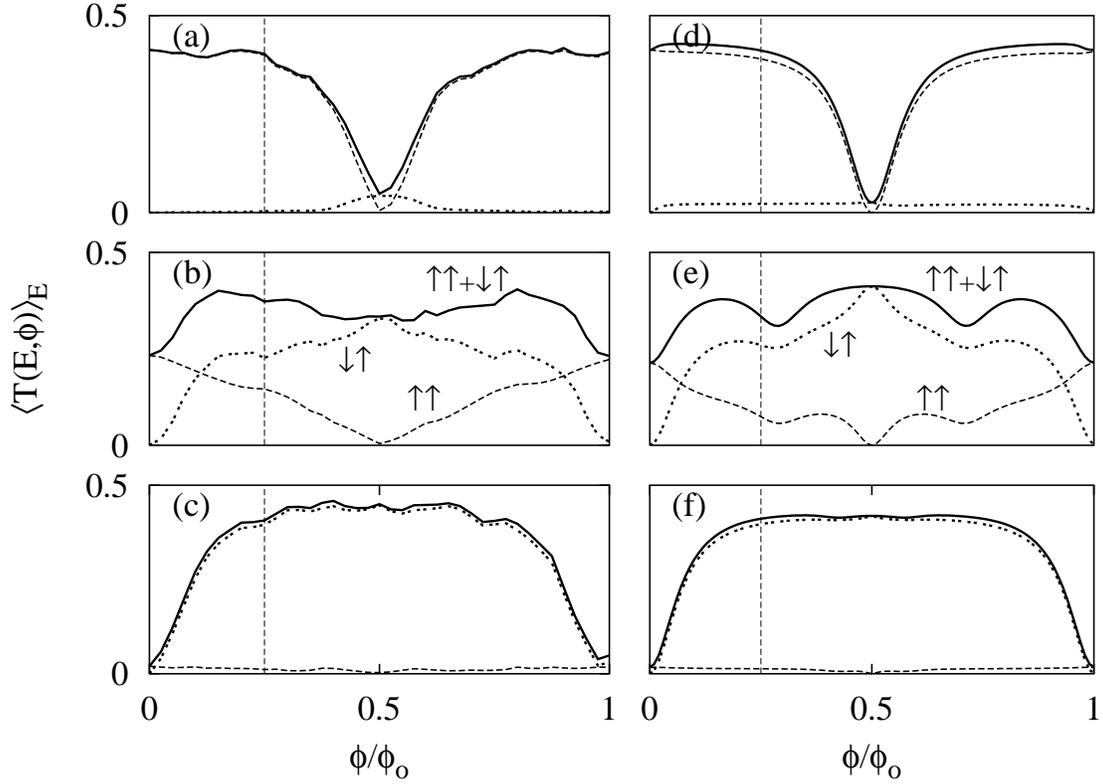,width=15cm,angle=-90}}
\end{center}
\vspace*{-5mm}

\caption{
Energy-averaged transmission of spin-up polarized electrons
(spin in $y$-direction, see text) through
a quasi-1D ring as function of a flux
$\phi=\pi r_0^2 B_0$
in the presence of a circular in-plane field $B_{\rm i} \gg B_0$
of increasing strength: (a) weak ($Q \approx 0.05$), (b) moderate
($Q \approx 0.7$),
(c) strong ($Q \approx 4$).
The overall transmission (solid line) is split into its components
$\langle T^{\uparrow \uparrow} \rangle$ (dashed) and
$\langle T^{\downarrow \uparrow}\rangle$ (dotted).
Note the change in the polarization upon tuning the flux
and particularly
the spin-switching mechanism at $\phi = \phi_0/2$.
Panels (d)-(f)
show corresponding calculations for a strictly 1D ring using
the transfer matrix approach (paper I) with coupling constant $\epsilon=0.3$.
These are to be compared with panels (a)-(c), respectively.
Equivalent results are
obtained for spin-down electrons (not shown here).
}
\label{Q1D-1D-s-switch}
\end{figure}

We consider transport of spin-up polarized electrons through a quasi-1D ring,
 Fig.~\ref{Q-1D-rings}(a), in the presence of the circular in-plane field
$\vec{B}_{\rm i}$ given by Eq.~(\ref{Bi}). The spin orientation is
defined with respect to the $y$-axis as shown in Fig.~\ref{Q-1D-rings}(b).
For the conductance calculations we apply the numerical
method outlined in Sec.~\ref{NA} and compare with 1D results from
paper I. To this end
we focus on the case where leads and ring support only one open
channel and choose the field configuration as quasi in-plane texture
where $B_0\! \ll\! B_{\rm i}$. The aspect ratio of the ring is  $d/r_0=0.25$.
Our numerical results for the energy-averaged transmission
$\langle T(E,\phi)\rangle_E$ are shown 
in Fig.~\ref{Q1D-1D-s-switch}(a)-(c) 
for three different scaled strengths
$Q=\omega_{\rm s}/\omega$ of the in-plane inhomogeneous field
($Q \ll 1;~Q \sim 1;~Q \gg 1$). 
In the weak-field limit ($Q \ll 1$,
Fig.~\ref{Q1D-1D-s-switch}(a)),
the electron spin is barely affected by the magnetic field. 
The total transmission (solid line) shows the usual AB oscillations and is
predominantly given by $\langle T^{\uarr\uarr}\rangle$ (dashed line), whereas
$\langle T^{\darr\uarr}\rangle$ (dotted line) is close to zero.
The behaviour is reversed in the adiabatic limit,
Fig.~\ref{Q1D-1D-s-switch}(c), 
where the spin-flip coefficient $\langle T^{\darr\uarr}\rangle$
carries the AB oscillations, now shifted by $\phi_0/2$ due
to the geometrical phase as discussed in Refs.~\onlinecite{FHR01,FR01},
and $\langle T^{\uarr\uarr}\rangle$ is, in turn, nearly zero.


Figure \ref{Q1D-1D-s-switch}(b) shows the general case of an intermediate
field ($Q \sim 1$). With increasing flux the polarization of transmitted
electrons
changes continuously. Most interestingly, $\langle T^{\darr\uarr} \rangle=0$
at $\phi = 0$, while $\langle T^{\uarr\uarr} \rangle = 0$ for
$\phi = \phi_0/2$.
This means that 
for zero flux an ensemble of
spin-polarized charge carriers is transmitted keeping
the incoming spin direction (spin-flip suppression),
while at $\phi =\phi_0/2$ the
transmitted electrons just reverse their spin direction.
In other words, by tuning the flux from $0$ to $\phi_0/2$, one
can reverse the polarization of transmitted particles in a controlled way.
Hence, the AB ring combined with the rotationally symmetric magnetic
field acts as a tunable spin switch. This mechanism proves to be {\em independent}
of the field strength $B_{\rm i}$ or $Q$,
which determines only the size of the spin-reversed current.
Alternatively, for a fixed flux $0 < \phi < \phi_0/2$
(vertical dashed line in Fig.~\ref{Q1D-1D-s-switch})
the spin polarization is reversed upon passing from the non-adiabatic to the
adiabatic regime, while the total transmission remains nearly constant.

The mechanism for changing the spin direction does
neither rely on the spin coupling to the control field $B_0$ (as
long as $B_0\! \ll\! B_{\rm i}$),
nor on the  Zeeman splitting often exploited in spin filters
\cite{GB00,PFR03b}. The effect exists for both spin-up and spin-down
polarized particles \cite{thesis}.
It is of pure quantum interference origin, due to a cooperation between charge-
(which couples to the flux $\phi$) and spin-
(which couples to the field $\vec{B}_{\rm i}$) coherence, and exists not only
for the smoothed energy-averaged transmission as shown in
Fig.~\ref{Q1D-1D-s-switch}
but also for the transmission at a given energy.

We further note that the spin-coupling to the inhomogeneous field produces an
attenuation of the AB oscillations in the overall transmission for moderate
fields, confirm Fig.~\ref{Q1D-1D-s-switch}(b). This is due to the comparable
amplitude of the $\phi_0/2$-shifted components $\langle T^{\uarr\uarr}
\rangle$ and $\langle T^{\darr\uarr} \rangle$.

Corresponding calculations for strictly 1D rings using the transfer
matrix approach (paper I) lead to similar results.
They are summarized in
Fig.~\ref{Q1D-1D-s-switch}(d)-(f), to be compared with the above
discussed results of Fig.~\ref{Q1D-1D-s-switch}(a)-(c), respectively.

%

\subsection{Necessary conditions for the spin-switching mechanism}
  \label{CSSM}

For a strictly 1D ring an analytical proof for the
spin switch effect is given in paper I. There we show rigorously
that the transmission coefficient $T^{\uarr\uarr}$ vanishes
completely at $\phi = \phi_0/2$, if
the magnetic field to which the spins couple
has no component perpendicular to the plane of the ring.
In the following we explore the range of validity of the spin-switch effect
discussed in Sec.~\ref{ABRSS}. To this end, we consider a set
of five alternative situations and generalizations with respect to the
previous case of a single-channel symmetric ring.


\begin{figure}
\begin{center}
\centerline{ \psfig{figure=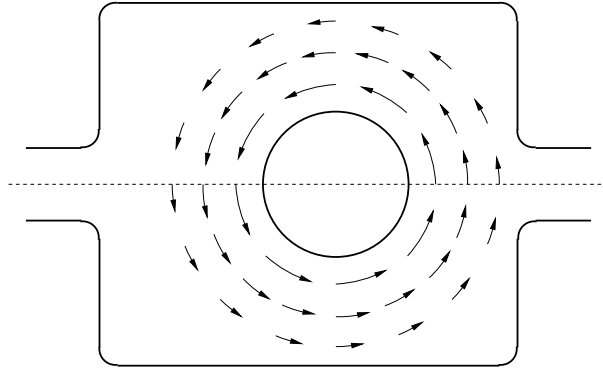,width=8cm,angle=-90}}
\end{center}
\vspace*{-5mm}

\caption{Ring-type geometry with a horizontal symmetry axis used for the
numerical calculations shown in Fig.~\ref{asymS-spin-switch}.
The circular in-plane field $\vec{B}_{\rm
i}$, Eq.~(\ref{Bi}), has its origin at the center of the inner disc.
}
\label{asymX-sinai}
\end{figure}

\begin{figure}
\begin{center}
\centerline{ \psfig{figure=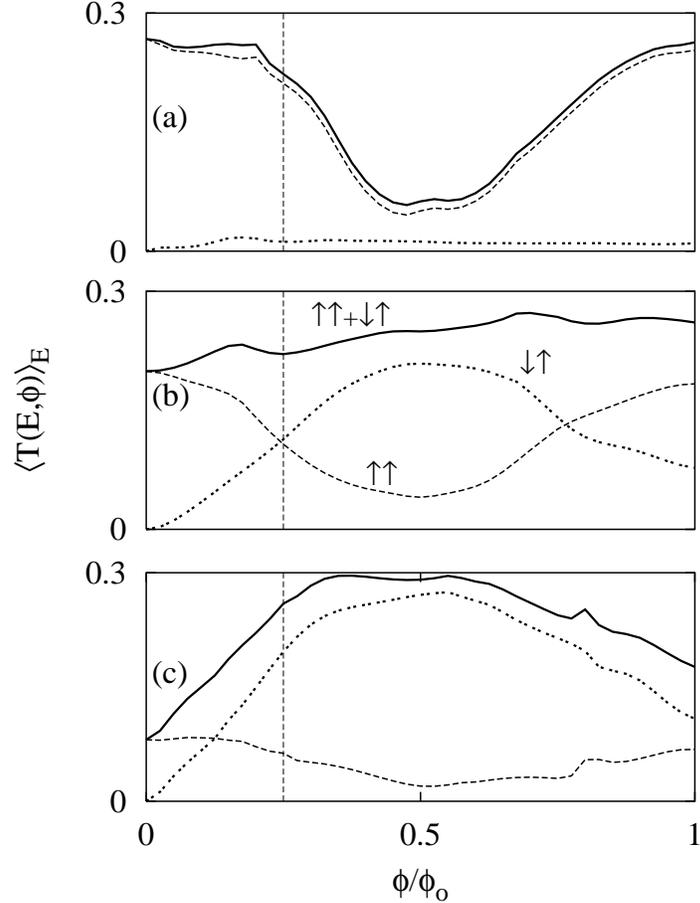,width=10cm,angle=0}}
\end{center}
\vspace*{-5mm}

\caption{Averaged transmission for spin-up polarized
incoming electrons through
the 2D ringwise geometry of Fig.~\ref{asymX-sinai} as function of the mean
flux $\phi=\pi r_0^2 B_0$ in the presence of a circular in-plane field $B_{\rm
i} \gg B_0$  of increasing strength: (a) weak, (b) moderate, (c) strong.
The overall transmission (solid line) is split into its components
$\langle T^{\uparrow \uparrow} \rangle$ (dashed) and
$\langle T^{\downarrow \uparrow}\rangle$ (dotted).
Note that the spin-switching mechanism at $\phi =
\phi_0/2$ is attenuated with respect to Fig.~\ref{Q1D-1D-s-switch} (due to
the large effective area of the inner cavity in Fig.~\ref{asymX-sinai}, see
text) but still present.
}
\label{asymS-spin-switch}
\end{figure}

{\it I)} To clarify whether the effect pertains for more general
geometries than a 1D ring, we study transport through a doubly-connected 2D
cavity,
a Sinai-type billiard as shown in Fig.~\ref{asymX-sinai}. This geometry still
obeys reflection symmetry with respect to the horizontal ($x-$) axis,
but not necessarily with respect to the vertical axis.
Numerical calculations are performed for leads supporting only one open
channel.
However, within the cavity the number of open modes is larger and
not well defined, since the system is not separable.
In Fig.~\ref{asymS-spin-switch} results are displayed for the energy-averaged
transmission $\langle T (E,\phi)\rangle_E$, again for incoming spin-up
polarized particles, 
in the presence of a weak (a),
moderate (b), and strong (c) inhomogeneous in-plane field.
The general features are similar
to those observed in the quasi-1D and 1D cases (Fig.~\ref{Q1D-1D-s-switch}).
However, some deviations appear. While the spin-flip suppression $\langle
T^{\downarrow \uparrow} \rangle = 0$ remains true for $\phi=0$, the component
$\langle T^{\uparrow \uparrow} \rangle$ does not vanish completely at
$\phi=\phi_0/2$, even though it still exhibits a pronounced minimum (panel (b)).
This is a consequence of the 
comparatively large fraction of the total magnetic flux
penetrating the accessible region of the Sinai cavity in
Fig.~\ref{asymX-sinai}, giving rise to a whole range of 'path-dependent'
accumulated fluxes. Consequently, the point $\phi=\phi_0/2$ is no longer
well defined.  We further note that, owing to the appearence of Berry phases
as in the quasi-1D case, the AB oscillations of  $\langle T \rangle$
in Fig.~\ref{asymS-spin-switch} (solid line) show again
a phase shift of $\phi_0/2$ near the adiabatic limit,
panel (c), with respect to the weak field limit, panel (a).

We conclude that despite relaxing the constraints of $(i)$ the vertical reflection
symmetry and $(ii)$ the quasi-1D nature of the ring geometry, the
spin-switch mechanism remains as an outstanding effect for a rather general class
of doubly-connected geometries.


\begin{figure}
\begin{center}
\centerline{ \psfig{figure=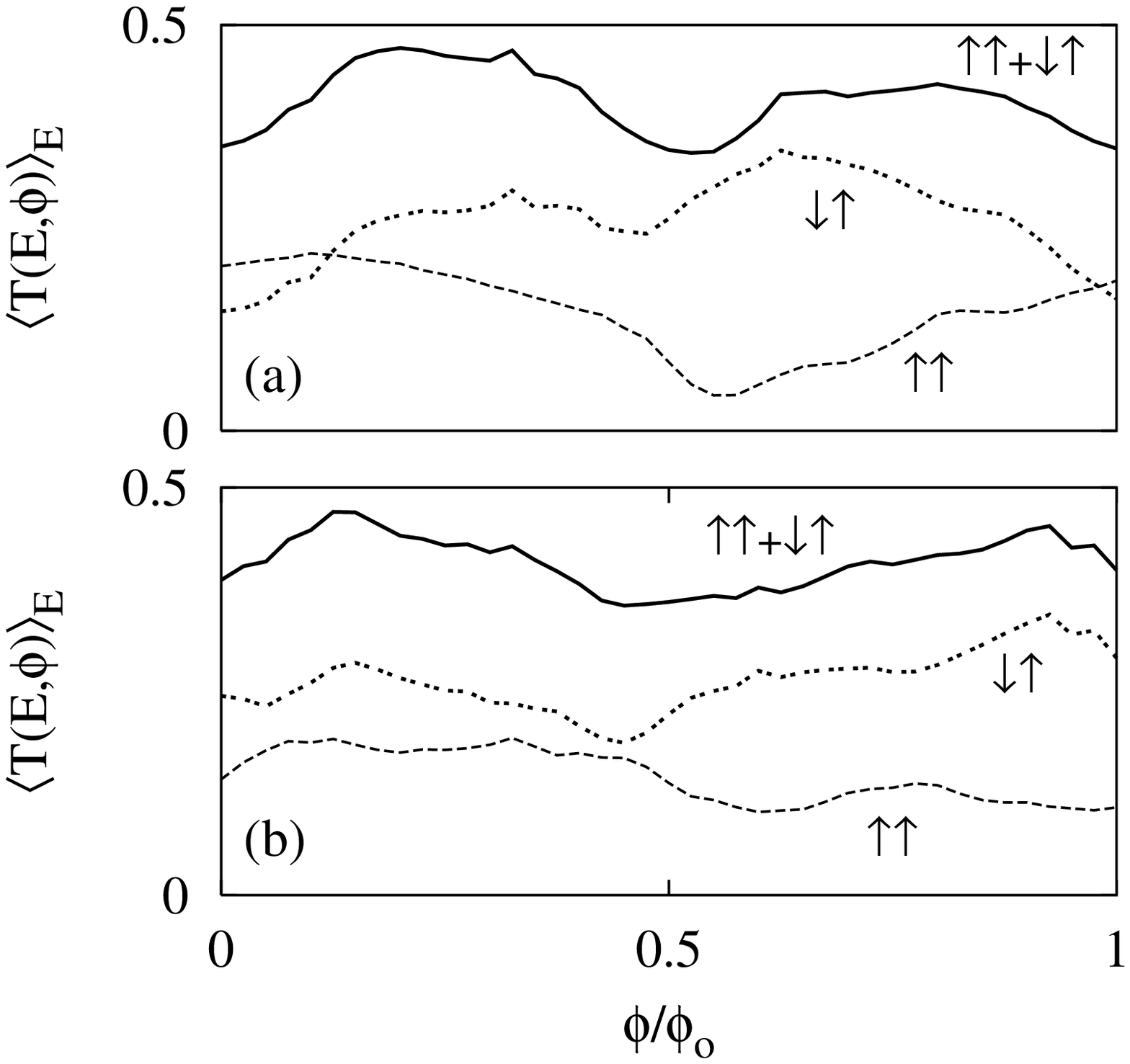,width=11cm,angle=0}}
\end{center}
\vspace*{-5mm}

\caption{Averaged transmission for spin-up polarized incoming
electrons through a quasi-1D ring with slightly asymmetric arms, see text.
The parameters defining the asymmetry via a vertical displacement of the inner
disk are (a) $\Delta_{\rm y}/d=0.05$
and (b) $\Delta_{\rm y}/d=0.10$. A moderate inhomogeneous field
strength, equivalent to that of Fig.~\ref{Q1D-1D-s-switch}(b), is
applied. Note the absence of the spin-switching effect in panel (b).
}
\label{asym.y.R-spin-switch}
\end{figure}

{\it II)} To investigate the role of the remaining horizontal reflection symmetry
we slightly shift the central disc of the symmetric ring in Fig.~\ref{Q-1D-rings}(a)
by an amount $\Delta_{\rm y}$ along the $y$-direction, while
keeping all other parameters as in Fig.~\ref{Q1D-1D-s-switch}(a)-(c).
This leads to a difference in the effective lengths of paths along the upper and lower
arms, breaking the horizontal reflection symmetry.
Numerical results for a moderate field strength (as in
Fig.~\ref{Q1D-1D-s-switch}(b)) are shown in
Fig.~\ref{asym.y.R-spin-switch}(a) and (b) for
a scaled shift $\Delta_{\rm y}/d=0.05$ and
$\Delta_{\rm y}/d=0.10$, respectively. In contrast to the symmetric case of
Fig.~\ref{Q1D-1D-s-switch}(b), both panels depict spin
flips ($\langle T^{\downarrow \uparrow} \rangle \neq 0$) already at
$\phi=0$. Nevertheless,
for the smaller asymmetry in 
panel (a) 
a distinct
modulation of the spin-dependent components $\langle
T^{\uparrow \uparrow}\rangle$ and $\langle T^{\downarrow \uparrow}\rangle$
as function of $\phi$ is observed, 
and a spin-reversed current $\langle
T^{\downarrow \uparrow}\rangle$ prevails near $\phi=\phi_0/2$.
With increasing deformation parameter $\Delta_{\rm y}$
the spin-switch effect is disappearing as
seen from  Fig.~\ref{asym.y.R-spin-switch}(b).

As a result we find that reflection symmetry with respect to the
axis defined by the two opposite leads is required for
spin-flip suppression at $\phi=0$ and spin-inversion at $\phi=\phi_0/2$.


\begin{figure}
\begin{center}
\centerline{ \psfig{figure=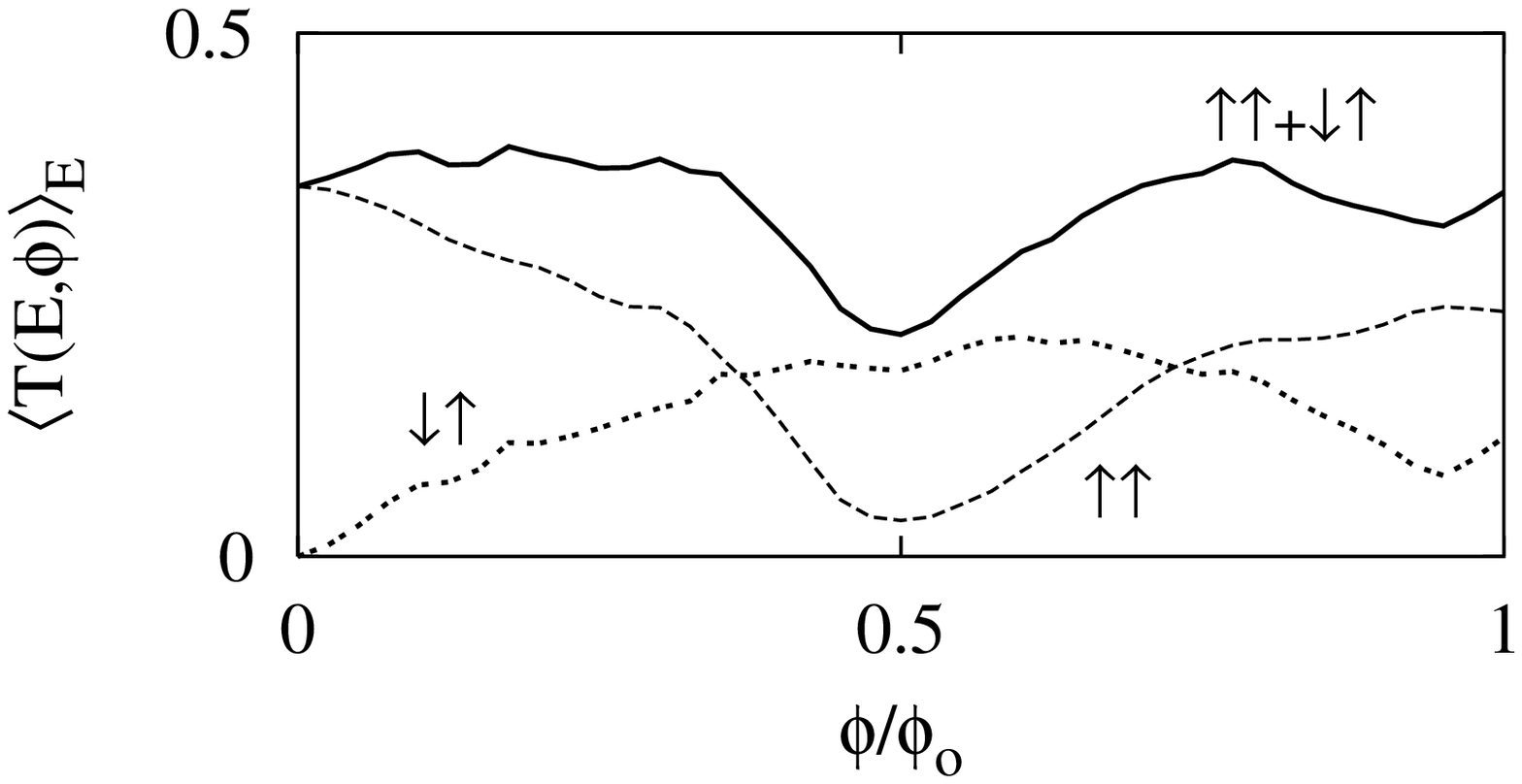,width=11cm,angle=0}}
\end{center}
\vspace*{-5mm}

\caption{Averaged transmission for geometry, field strength and asymptotic
spin-states equivalent to those of Fig.~\ref{Q1D-1D-s-switch}(b), this
time
for comparable fields $B_{\rm i} \sim B_0$ at $\phi=\phi_0/2$ (field
texture displayed in Fig.~\ref{crown-field}). The spin-coupling with the
field $B_0$ generating the flux is not negligible.
}
\label{symR-spin-switch-cone}
\end{figure}


{\it III)} In the following we
return to the symmetric ring, but 
relax the constraint $B_0 \ll B_{\rm i}$
of the in-plane field configuration.
We consider an inhomogeneous field $B_{\rm i}$
of moderate strength such that $B_{\rm i} \sim B_0$ at $\phi=\phi_0/2$. As a
consequence, the overall field texture is characterized by a finite
tilt angle $\alpha \neq \pi/2$ as displayed in
Fig.~\ref{crown-field}. Numerical results for single-channel transmission
of incoming spin-up particles
are shown in Fig.~\ref{symR-spin-switch-cone}.
Although no longer perfect, 
spin-switching still takes place near
$\phi=\phi_0/2$. Thus, the spin-coupling to
the field generating the flux which is not negligible in this case,
produces an attenuation of the effect
with respect to in-plane field result; the overall
effect, however, still persists. This is similar to
the result found in
Fig.~\ref{asymS-spin-switch}(b) for a ring-type geometry with large aspect
ratio (Fig.~\ref{asymX-sinai}).
In the remaining part of the paper we return to the situation of a
quasi in-plane field, $B_0 \ll B_{\rm i}$.


\begin{figure}
\begin{center}
\centerline{ \psfig{figure=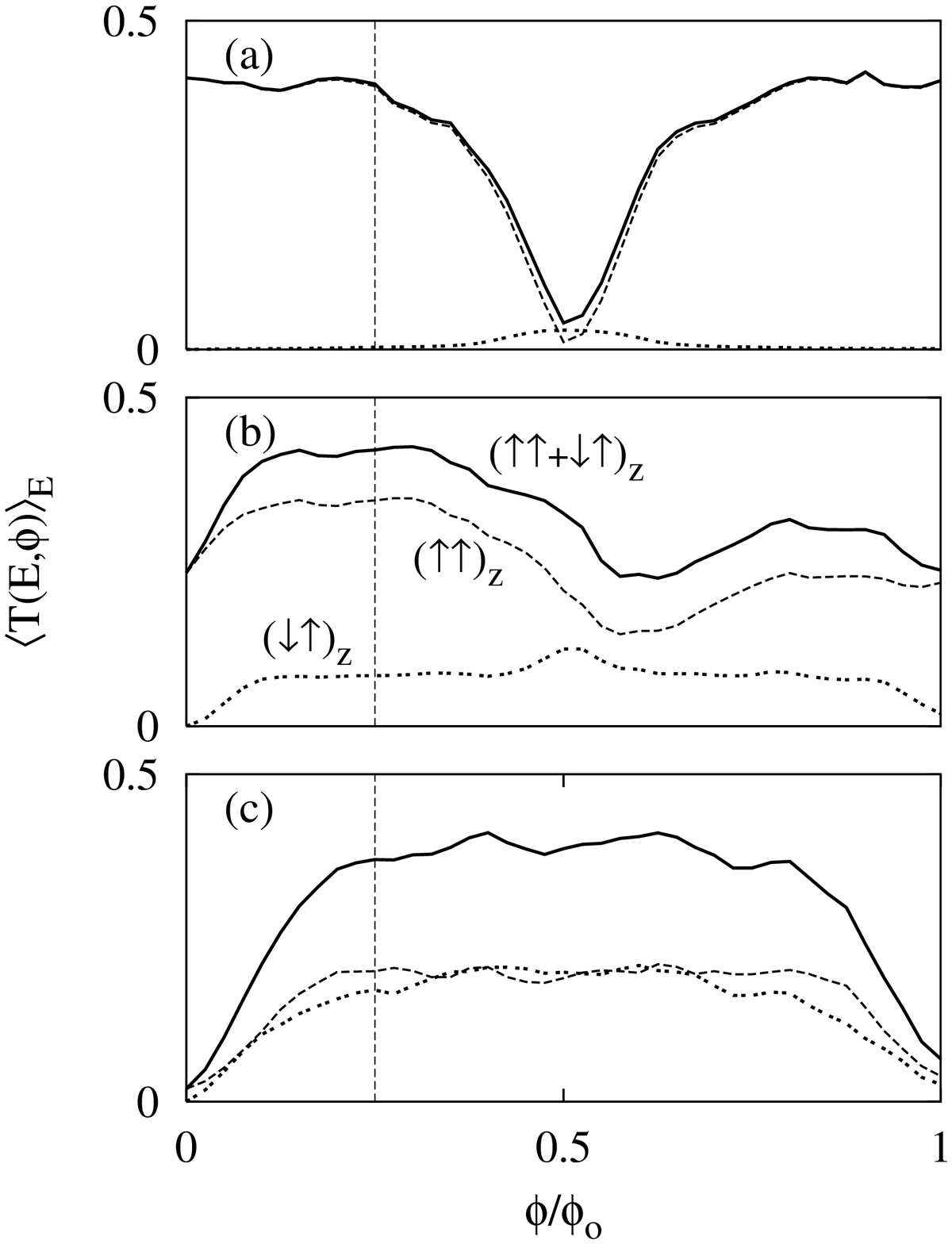,width=10cm,angle=0}}
\end{center}
\vspace*{-5mm}

\caption{Averaged transmission in the in-plane case ($B_{\rm i} \gg B_0$), 
geometry and field strengths 
equivalent to those of Fig.~\ref{Q1D-1D-s-switch}(a)-(c), but
for asymptotic spin-states defined with respect to the $z$-axis.
}
\label{symR-spin-switch-Z}
\end{figure}


{\it IV)} So far the spin-switching mechanism has been considered
for asymptotic spin states polarized in the in-plane
$\pm y$-direction, i.e. for eigenvectors of the Pauli matrix $\sigma_{\rm y}$.
In order to see whether the effect depends on the polarization direction
with respect to the field texture, we introduce asymptotic
spin-states orientated along the $z$-axis perpendicular to the 2D system,
i.e.\  eigenvectors of $\sigma_{\rm z}$.
The transmission amplitudes between $z$-orientated asymptotic
spin-states, $t^{s'_{\rm z}s_{\rm z}}$, are a linear combination of those
defined along the $y$-axis, $t^{s'_{\rm y}s_{\rm y}}$.
Numerical results for corresponding spin-up
electrons are displayed in Fig.~\ref{symR-spin-switch-Z}
for a symmetric quasi-1D ring in the presence of a weak (a), moderate (b),
and strong (c) in-plane inhomogeneous field (same parameters as in
Fig.~\ref{Q1D-1D-s-switch}(a)-(c)).
In the weak-field limit, Fig.~\ref{symR-spin-switch-Z}(a), the component
$\langle T^{(\darr\uarr)_{\rm z}}\rangle$ (dotted line) is close to zero, and
the total transmission (solid line) shows AB oscillations predominantly
given by $\langle T^{(\uarr\uarr)_{\rm z}}\rangle$ (dashed line) as expected.
In the opposite strong-field (adiabatic) 
limit, Fig.~\ref{symR-spin-switch-Z}(c), the AB
oscillations exhibit a phase shift of $\phi_0/2$ due to geometrical
phases. The spin polarization of transmitted particles, however, is not
reversed
as in the case of incoming
$y$-polarized spins (Fig.~\ref{Q1D-1D-s-switch}(c)). On the contrary, the
coefficients $\langle T^{(\uarr\uarr)_{\rm z}}\rangle$ and $\langle
T^{(\darr\uarr)_{\rm z}}\rangle$ have similar phase and magnitude,
leading to randomization of the orientation of transmitted spins.
This is a consequence of the spin precession
taking place around the local field direction during transport in the
adiabatic limit \cite{noteZad}.
For moderate field strengths, Fig.~\ref{symR-spin-switch-Z}(b), no
pronounced spin-switching mechanism appears either, though the spin-reversed
component
$\langle T^{(\darr\uarr)_{\rm z}}\rangle$ is maximum at $\phi=\phi_0/2$.
A spin-flip suppression remains at $\phi=0$ independent of $B_{\rm i}$.

\begin{figure}
\begin{center}
\centerline{ \psfig{figure=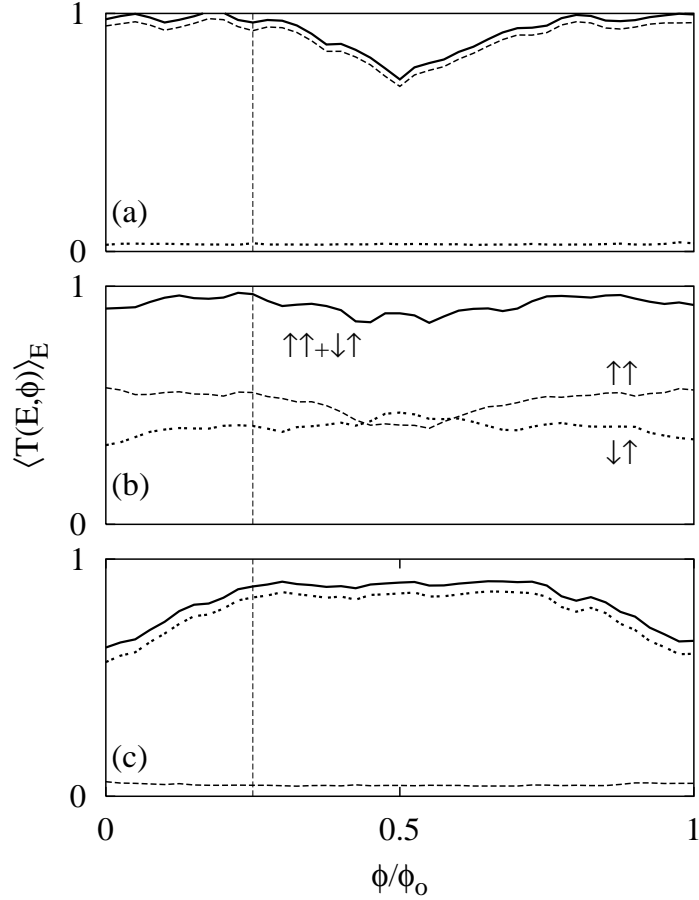,width=10cm,angle=0}}
\end{center}
\vspace*{-5mm}

\caption{Multichannel averaged transmission
as function of a flux $\phi=\pi r_0^2 B_0$ for
incoming electrons in spin-up states (oriented along the $y$-direction)
traversing a symmetric ring structure (Fig.~\ref{Q-1D-rings}(a))
in the presence of a circular in-plane field $B_{\rm i} \gg B_0$
of increasing strength: (a) weak, (b) moderate, (c) strong. The overall
transmission (solid line) is split into its components $\langle T^{\uarr
\uarr}\rangle = \sum_{n,m}^N \langle T^{\uarr \uarr}_{nm}\rangle$ (dashed) and
$\langle T^{\darr \uarr}\rangle=\sum_{n,m}^N \langle T^{\darr
\uarr}_{nm}\rangle$ (dotted) with $N=2$ (see text).
}
\label{symR-spin-switch-CH2}
\end{figure}

{\it V)} Finally we consider the case of {\em multichannel}
transport, namely leads of width $w$ supporting more than one open channel
($N={\rm Int}[k_{\rm F}w/\pi]>1$). Consequently, the spin-dependent
transmission coefficients $T^{s's}$ consist now of a sum over incoming ($m$)
and outgoing (n) modes
$T^{s's}=\sum_{n,m}^N T^{s's}_{nm}$, see Eq.~(\ref{landa}).
Without loss of generality, we discuss here the case $N=2$ and return to our
earlier definition of asymptotic spin-states as eigenvectors of $\sigma_{\rm y}$.
Fig.~\ref{symR-spin-switch-CH2} shows numerical results for the total
transmission of incoming spin-up polarized charge carriers through a symmetric ring
as illustrated in 
Fig.~\ref{Q-1D-rings}(a). (The geometry parameters are
equivalent to those used for the calculations in
Fig.~\ref{Q1D-1D-s-switch}(a)-(c); the energy-average is taken
between the (open) second and (still closed) third channel. 
The different panels correspond to a
weak (a), moderate (b), and strong (c) inhomogeneous in-plane field.
As expected, the main contribution to the total transmission
(solid lines) in the weak field limit, given by $\langle
T^{\uarr\uarr}\rangle$ (dashed line in panel (a)), is replaced by $\langle
T^{\darr\uarr}\rangle$ (dotted line in panel (c)) when the adiabatic
limit is approached. Simultaneously, the AB oscillations are shifted due to
geometrical phases.
Furthermore, we no longer observe both spin-flip suppression at $\phi=0$
($\langle T^{\darr\uarr}\rangle \neq 0$) and polarization inversion at
$\phi=\phi_0/2$ ($\langle T^{\uarr\uarr}\rangle \neq 0$) for  moderate field
strengths in panel (b).


\begin{figure}
\begin{center}
\centerline{ \psfig{figure=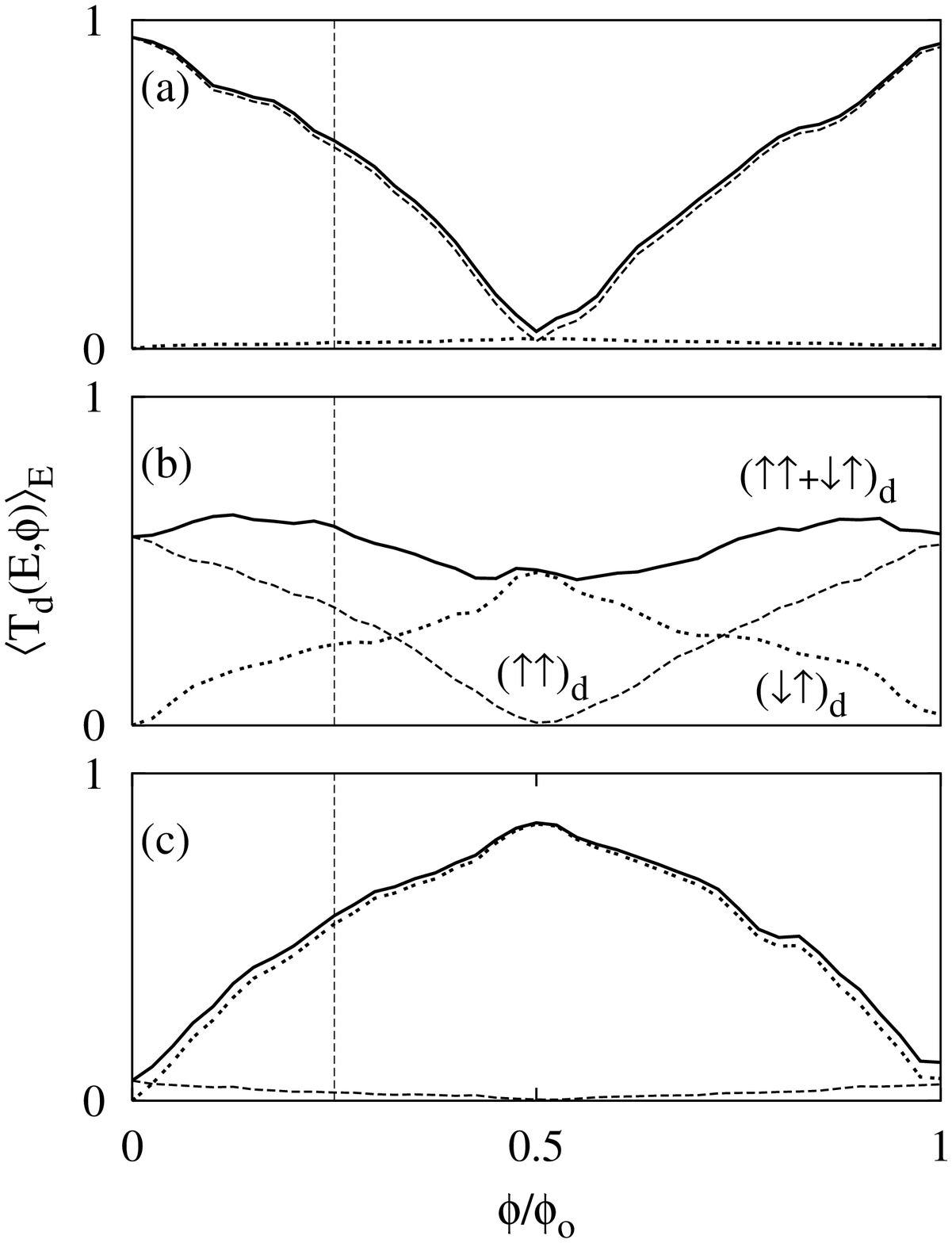,width=10cm,angle=0}}
\end{center}
\vspace*{-5mm}

\caption{Diagonal contribution (in mode number) to the multichannel
averaged tranmission of Fig.~\ref{symR-spin-switch-CH2}. The overall diagonal
transmission $\langle T_{\rm d}\rangle$ (solid line) is split into its components
$\langle T^{\uarr \uarr}_{\rm d}\rangle = \langle T^{\uarr \uarr}_{11}\rangle
+ \langle T^{\uarr \uarr}_{22}\rangle$
(dashed) and $\langle T^{\darr \uarr}_{\rm d}\rangle = \langle T^{\darr
\uarr}_{11}\rangle + \langle T^{\darr \uarr}_{22}\rangle$ (dotted). Note the
similarity with the results of Fig.~\ref{Q1D-1D-s-switch}.
}
\label{symR-spin-switch-CH2-d}
\end{figure}

\begin{figure}
\begin{center}
\centerline{ \psfig{figure=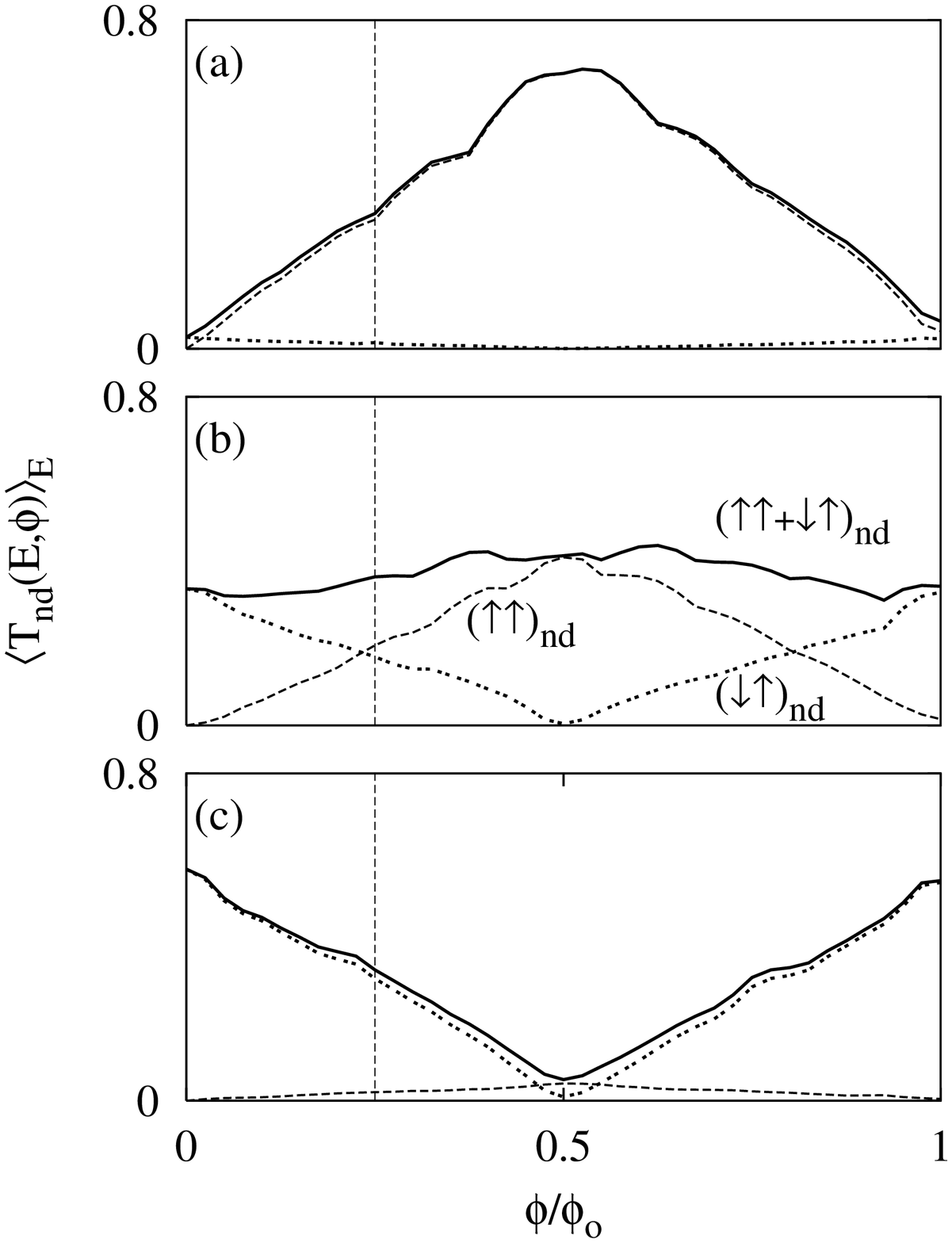,width=10cm,angle=0}}
\end{center}
\vspace*{-5mm}

\caption{Off-diagonal contribution (in mode number) to the multichannel
averaged tranmission of Fig.~\ref{symR-spin-switch-CH2}. The overall
off-diagonal transmission $\langle T_{\rm nd}\rangle$ (solid line) is split
into
its components $\langle T^{\uarr \uarr}_{\rm nd}\rangle = \langle T^{\uarr
\uarr}_{12}\rangle + \langle T^{\uarr \uarr}_{21}\rangle$
(dashed) and $\langle T^{\darr \uarr}_{\rm nd}\rangle = \langle T^{\darr
\uarr}_{12}\rangle + \langle T^{\darr \uarr}_{21}\rangle$ (dotted).
Note the contrast with the complementary diagonal contribution in
Fig.~\ref{symR-spin-switch-CH2-d}.
}
\label{symR-spin-switch-CH2-nd}
\end{figure}

Additional
insight is gained by analyzing the respective contributions of the
different current carrying transverse modes in the leads.
Consider the transmission coefficients diagonal and non-diagonal in
channel number, namely, $T^{s's}_{\rm d} \equiv T^{s's}_{11}+T^{s's}_{22}$ and
$T^{s's}_{\rm nd} \equiv T^{s's}_{12}+T^{s's}_{21}$. The corresponding
energy-averaged quantities are shown in Figs.~\ref{symR-spin-switch-CH2-d} and
\ref{symR-spin-switch-CH2-nd}, respectively. The results are organized as
in Fig.~\ref{symR-spin-switch-CH2}. The traces in
Fig.~\ref{symR-spin-switch-CH2-d} show on the one hand
that the diagonal term $\langle T_{\rm d}\rangle$ exhibits the same
features as for single-channel transport, see Fig.~\ref{Q1D-1D-s-switch},
including the spin-switch mechanism. On the other hand,
we observe precisely the opposite behaviour for the off-diagonal
component $\langle
T_{\rm nd}\rangle$ in Fig.~\ref{symR-spin-switch-CH2-nd}.
This implies for zero flux that an ensemble of incoming
polarized spins within, e.g., the first channel $m=1$ is partially
transmitted keeping the original spin polarization only within the first
outgoing channel $n=1$. At the same time, a complementary spin-reversed
fraction of particles is leaving the cavity  through the second outgoing channel,
$n=2$. For  flux $\phi=\phi_0/2$,
the spin-reversed fraction exits the system through the lowest mode $n=1$,
while the spins keep the original orientation within the second mode, $n=2$.
In other words, by tuning $\phi$ from 0 to $\phi_0/2$ one can control,
although not independently,
the spin polarization of each outgoing channel provided that the
incoming electrons were spin-polarized.
Furthermore, we point out that for $\phi=0$ the
spin-coupling to a finite in-plane field leads to non-zero
(odd) cross terms in channel number. This is remarkable, since such cross
terms do not contribute to the transmission of
particles without spin for systems
with horizontal reflection symmetry
(see e.g.\ the weak field limit of
Fig.~\ref{symR-spin-switch-CH2-nd}(a) at $\phi=0$).


We close by mentioning that the amplitude 
of the AB oscillations
in the total transmission for weak fields, solid line in
Fig.~\ref{symR-spin-switch-CH2}(a), is relatively small compared with its
diagonal component, Fig.~\ref{symR-spin-switch-CH2-d}(a).
This reflects the counteracting r\^ole played by the off-diagonal contribution
(Fig.~\ref{symR-spin-switch-CH2-nd}(a)) in the total transmission.


%
\section{Summary and conclusion}

We have studied {\it non-adiabatic} spin-dependent transport through
ballistic conductors of different shape (straight and ring-type
geometries) subject to inhomogeneous magnetic fields of varying
strength.
Our account generalizes studies of the regime of adiabatic spin-transport,
widely discussed in the literature \cite{Ber84,Ste92,FR01,LGB90}. This
regime is included here as the strong field limit.

For straight conductors we discussed several spin effects in the
quantized conductance. In particular we found a strong enhancement
of the adiabatic spin channel each time a new transverse mode opens in the
conductor, owing to the fact that electrons propagate slowly within the channel
corresponding to the new mode.

For ring geometries we obtain numerically the explicit dependence
of the transmission on the scaled field strength $Q$, which acts as an
adiabaticity parameter, elucidating the r\^ole of geometrical phases
in ballistic quantum transport and possible experimental realizations.
Moreover, for in-plane field configurations and symmetric ballistic ring
microstructures we
demonstrate how an additional small flux $\phi$ can be used to control the spin
dynamics and thereby tune the polarization of transmitted electrons \cite{FHR01}.
This quantum mechanism, which is analytically investigated in detail
in a subsequent paper \cite{HSFR03} does not require adiabaticity.
We have also assessed in detail the range of validity of the spin-switch effect
for various different situations relaxing constraints on symmetry, field
configuration, and channel number.
In combination with a spin detector such a device may be used to control spin
polarized current, similar to the spin field-effect transistor
proposed in Ref.~\onlinecite{DD90}.
For metallic, generally diffusive conductors disorder breaks the spatial
symmetry. We found numerically that the spin switch mechanism no longer
prevails for diffusive rings \cite{Pop02}.

Finally, we point out that ballistic rings with Rashba (spin-orbit)
interaction \cite{noteNMT99}, yielding an effective in-plane magnetic field
in the presence of a vertical electric field, exhibit a similar
spin-switch effect \cite{FR02}.

\acknowledgments
A larger part of this work was performed at the
{\em Max-Planck-Institut for the Physics of Complex Systems} in
Dresden, Germany. We thank the institute
and particularly P.~Fulde for continuous support.
We gratefully acknowledge financial support from the {\em Alexander von
Humboldt Foundation} and the {\em Deutsche Forschungsgemeinschaft}
through the research group {Ferromagnet-Semiconductor Nanostructures}.
We also thank J.~Fabian, H.~Schomerus, and D.\ Weiss for many
useful discussions.



\begin{references}
\bibitem{noteFR01} Exceptions are e.g.\ the numerical study of
Berry phase effects in adiabatic spin-dependent transport \cite{FHR01,FR01}
and Refs.~\onlinecite{NF01,PB02} addressing spin-orbit effects in quantum transport.
\bibitem{noteGauge} The choice $\vec{A}_{\rm em}= B_{\rm
i}~\Theta(x)~\Theta(L-x)~[z~(1+ (\pi/L) y)~\cos[(\pi/L) x]~\hat{x} +
y~\sin[(\pi/L) x]~\hat{z}]$
generates the in-plane field at $z=0$ introduced in Fig.~\ref{2D-lead}.
The $\delta$-function type
fields arising from differentiation of the Heaviside step functions
$\Theta$ cancel due to the vanishing $z$-component of $\vec{A}_{\rm em}$ at
$x=0$ and $x=L$. We further note that $\vec{\nabla} \cdot \vec{A}_{\rm em}=0$
for $z=0$.
\bibitem{noteModes} The $n$th mode propagates along the $x$-direction with
velocity $v_x(n)=(v_{\rm F}^2-v_n^2)^{1/2}$, where $v_{\rm F}$ is the Fermi
velocity, $v_n=(\hbar/m^*)k_n$ is the velocity in
$y$-direction and $k_n=(\pi/w)n$ is the corresponding quantized wave number.
\bibitem{noteKABMR00} Such fields, induced by currents through
$\sim$100 nm narrow pillars, were realized in the context
of Oerstedt switching \cite{KABMR00}. With current densities of order
$10^7$ A/cm$^2$ one can achieve field strengths of a few hundred mT.
\end{references}
\end{document}